\documentclass[preprint,amsmath,nofootinbib,superscriptaddress]{revtex4}
\usepackage{setspace}
\usepackage{graphicx}
\usepackage{bm}
\usepackage{epsfig}
\usepackage{color}
\usepackage{colordvi}
\newcommand{\nn}{\nonumber}

\newcommand{\beq}{\begin{equation}}
\newcommand{\eeq}{\end{equation}}
\newcommand{\bea}{\begin{eqnarray}}
\newcommand{\eea}{\end{eqnarray}}

\begin{document}

\title{Next to Leading Order Spin(1)Spin(1) Effects in the Motion of Inspiralling Compact Binaries} 

\author{Rafael A. Porto} 
\affiliation{Department of Physics, University of California Santa Barbara, CA 93106}
\author{Ira Z. Rothstein}
\affiliation{Department of Physics, Carnegie Mellon University,
Pittsburgh, PA 15213}
\begin{abstract}
Using effective field theory techniques we compute the next to leading order Spin(1)Spin(1) terms in the potential of spinning compact objects at third Post-Newtonian (PN) order, including subleading self-induced finite size effects. This result represents the last ingredient to complete the relevant spin potentials to 3PN order from which the equations of motion follow via a canonical formalism. As an example we include the precession equation.
\end{abstract}

\maketitle

\section{Introduction}
There has been significant progress in calculating higher order corrections to the potentials
for spinning compact binaries. The spin potentials to 3PN order, from which the ${\cal O}({\bf S}_1{\bf S}_2)$ contribution to the  equations of motion (EOM) follow via a canonical procedure, were reported in \cite{eih} for direct spin--spin interactions, and in \cite{comment} from indirect spin--orbit effects. These results were presented using the Newton-Wigner (NW) spin supplementarity condition (SSC) at the level of the action. The latter procedure was shown to be accurate up to 4PN order in \cite{nrgr5} using standard power counting techniques.  
These results were derived using the NRGR formalism, which appears to provide simple tools to compute in the PN expansion \cite{nrgr,nrgr1,nrgr2,nrgr3,nrgr4,nrgr5,nrgr6}, as well as the extremal limit \cite{workadam,chad}. More recently we computed the  spin potentials to 3PN order in the covariant SSC \cite{ss}, and explicitly showed the equivalence with our previous results in \cite{eih,comment}. 
The computations using the NW SSC at the level of the action were also re-derived within the NRGR  approach in \cite{levi}, albeit using a different choice of metric parameterizations introduced in \cite{kol2}.  A recent calculation of the ${\bf S}_1{\bf S}_2$ Hamiltonian to 3PN appeared in \cite{Schafer3pn}. The results in \cite{Schafer3pn} helped to clarify the necessity of taking into account spin--orbit effects to compute the ${\cal O}({\bf S}_1{\bf S}_2)$ contributions in the EOM to 3PN when one is working with the SSC at the level of the action. The equivalence between ours, and the more traditional approach of \cite{Schafer3pn}, was shown in \cite{comment,ss}. 

In \cite{ss} we worked within a Routhian formalism originally introduced in \cite{yee}, and developed in \cite{nrgr5,ss} within NRGR, which incorporates the (covariant) SSC and its conservation upon evolution in a canonical framework to all orders. Within the formalism of \cite{nrgr5,ss} the subleading spin--orbit effect in the EOM is proportional to $S^{j0}$, the spin tensor in the local frame, and contribute at ${\cal O}({\bf S}_1{\bf S}_2)$ after one reduces spin to a three vector using the SSC and takes into account the transformation between the local and global PN frames. Ultimately, the equivalence between the independent results computed in \cite{eih,comment} and \cite{ss}, with those in \cite{Schafer3pn} using ADM techniques \cite{adm}, confirms the validity of the new results at 3PN.\\

In \cite{ss} we sketched the necessary steps towards computing the next to leading order (NLO) ${\cal O}({\bf S}_i^2)$  contributions to the potential, which is the subject of this  paper. We first review the Routhian approach, emphasizing the inclusion of higher dimensional operators and the preservation  of the SSC
under time evolution.  We break up the calculation into three distinct types of contributions: 
 The  non--linear interactions, the finite size effects, and the contribution from the extra piece of the Routhian of \cite{ss}, which must be included to insure that the  SSC is preserved upon evolution. As an example we include the precession equation to 3PN. In appendix B we show how our result reduces to the geodesic motion around a Kerr background in the extreme mass ratio limit. 
 
\section{NRGR and spin effects in the Routhian approach}

One can introduce a Routhian to describe spin dynamics in a gravitational field as follows \cite{nrgr5,ss}
\begin{equation} \label{actR}
{\cal R} =-\sum_q \left( m_q \sqrt{u^2_q} + 
\frac{1}{2}S_{q}^{ab}\omega_{ab\mu} u^\mu_q +\frac{1}{2m_q}R_{d
e a b}(x_q)S^{c d}_{q} S^{a b}_{q} \frac{u^e_q u_{qc}}{\sqrt{u^2}}
+\ldots\right),
\end{equation}
where the ellipses represent non--linear terms in the curvature necessary to account for the mismatch between $p$ and $u$ once the SSC is enforced. Since at 3PN order we can consider the covariant SSC to be $S^{ab}u_b=0$, the higher order terms are irrelevant for our purposes. There is of course no obstruction to include higher order effects.
The overall minus sign is  chosen to ensure the spinless Feynman rules are not modified and the equations of motion (EOM) follow from
\begin{equation}
\frac{\delta }{\delta x^\mu}\int {\cal R} d\lambda=0, \;\;\; \frac{d
S^{ab}}{d\lambda} = \{S^{ab},{\cal R}\}, \;\;\; \frac{d
S^{ab}}{d\lambda} = \{V, S^{ab}\}\label {eomV},
\end{equation}
where  the potential is given by $V= -{\cal R}$, and 
\begin{eqnarray}
\{x^\mu ,{\cal P}_\alpha\}&=& \delta^\mu_\alpha,\;\;\; \{x^\mu,p_\alpha\}=\delta^\mu_\alpha,  \;\;\; \{{\cal P}^\alpha,{\cal P}^\beta\}= 0, \\
 \{x^\mu,x^\nu\} &=& 0, \;\;\; \{p^\alpha,p^\beta\} = \frac{1}{2}{R^{\alpha\beta}}_{ab}S^{ab}, \label{pp}\\
\{x^\mu ,S^{ab}\}&=& 0, \;\;\; \{p_\alpha,S^{ab}\}= \omega^{c[a}_\alpha S^{b]c} , \;\;\; \{{\cal P}^\alpha,S^{ab}\}= 0 \label{ps} \\
\{S^{ab},S^{cd}\} &=& \eta^{ac} S^{bd}
+\eta^{bd}S^{ac}-\eta^{ad} S^{bc}-\eta^{bc}
S^{ad} \label{als}
\end{eqnarray}
with $p^\mu$ given by $ {\cal
P}^\mu = p^\mu + \frac{1}{2}\omega^\mu_{ab} S^{ab}$, and ${\cal P}^\mu$ the canonical momentum. It is easy to show the Mathisson-Papapetrou equations \cite{papa} follow from (\ref{eomV}), and the Riemann dependent term in the Routhian guarantees the SSC is preserved upon evolution \cite{ss}.\\

Notice that our Routhian is similar to the one introduced in \cite{yee}, after the replacement
\beq
\frac{1}{2m_q}R_{d
e a b} S^{c d} S^{a b} \frac{u^e u_{c}}{\sqrt{u^2}} \to \frac{1}{m_q}\frac{Dp^d}{d\lambda}\frac{S^{dc}u_c}{\sqrt{u^2}}\label{dsu}.
\eeq

The replacement in (\ref{dsu}) follow from a change of variables.
Furthermore, 
  one can show that the net effect of this change 
\cite{yee} modifies the spin-gravity coupling as follows \footnote{In addition to this change one also generates higher dimensional operators which lead to effects which are beyond our interest in this
paper.}
\beq
-\frac{1}{2}\omega_\mu^{ab} {S}_{ab} u^\mu\rightarrow -\frac{1}{2}\omega_\mu^{ab} {\cal S}_{ab} u^\mu \label{newcals},
\eeq
where
\beq
{\cal S}^{ab} = S^{ab} + \frac{u_c}{u^2}S^{c[a}u^{b]}\label{newsc},
\eeq
or equivalently ($q=1,2$)
\beq
{\cal S}_q^{i0} = S_q^{ij}v_q^j  + \ldots
\label{caso}
\eeq
and
\beq
\label{calsrep}
{\cal S}^{ij}_q = S^{ij}_q + (S^{0i}_q-v_q^kS_q^{ki})v_q^j - (S_q^{0j}-v_q^kS_q^{kj})v_q^i +\ldots .
\eeq

From here it easy to show that, when the Routhian is {\it equivalently} written in terms of $\cal S$ using (\ref{dsu}), the covariant SSC is conserved, since
\beq
\label{cov0}
\frac{d}{dt} (S^{ab}u_b)= u_b \{S^{ab}, {\cal R}_0 ({\cal S}^{ab})\}+{\dot u}_d\frac{u_c}{u^2}\{ S^{ab},S^{cd}\}u_b  + S^{ab}{\dot u}_b = 0,
\eeq
where 
\beq
{\cal R}_0=-\sum_q \left( m_q \sqrt{u^2_q} + 
\frac{1}{2}{\cal S}_{q}^{ab}\omega_{ab\mu} u^\mu_q\right) .
\eeq
The expression in (\ref{cov0}) follows from the identity
\beq
u_b \{S^{ab}, {\cal R}_0 ({\cal S}^{ab})\}=0\label{ros},
\eeq
due to the fact that the spin algebra in terms of ${\cal S}^{ab}$ modifies the expression in (\ref{als}) by shifting $\eta^{ab} \to \eta^{ab} - \frac{u^au^b}{u^2}$. 
The inclusion of finite size effects obviously modify the EOM and the constancy of the SSC upon evolution is not guaranteed. However, from (\ref{cov0}) and (\ref{ros}) it is clear that  higher dimensional operators describing the internal structure of the bodies that are written  in terms of ${\cal S}^{ab}$ will preserve the consistency of the covariant SSC as shown in \cite{yee}. 

In the NRGR formalism, {\it physically relevant } higher dimensional operators are those which are written in terms of the electric, $E_{ab}$, and magnetic, $B_{ab}$, components of the Weyl tensor \cite{nrgr,nrgr2,nrgr3,nrgr6}. Terms which are proportional to the Ricci tensor, or scalar, can be removed by a field redefinition since they vanish on--shell \cite{nrgr}. As we mentioned earlier, in order to preserve the  SSC constraint upon evolution one needs to use ${\cal S}^{ab}$ for the higher dimensional terms in the wordline action. However, once these terms are written as a functions of the electric and magnetic part of the Weyl tensor, it is easy to show that using $S^{ab}$ still preserves the SSC,  as a consequence of the SSC itself and the orthogonality relation, $E_{ab}u^b=B_{ab}u^b=0$.  For example, the first higher dimensional operator we encounter is the self--induced quadrupole--like term, which written in terms of ${\cal S}^{ab}$ takes the form \cite{eih,nrgr3,ss} ($q=1,2$) 
\begin{equation}
L_{E{\cal S}^2} = \frac{C^{(q)}_{ES^2}}{2m m_p}\frac{E_{ab}}{\sqrt{u^2}}
{{{\cal S}}^a}_c {\cal S}^{cb}.\label{s2nc} 
\end{equation}  

If we now take $L_{E{\cal S}^2}$ and expand ${\cal S}^{ab}$ in terms of $S^{ab}$ using (\ref{newsc}) it is easy to show that the difference between (\ref{s2nc}) and
\begin{equation}
L_{ES^2} = \frac{C_{ES^2}}{2m m_p}\frac{E_{ab}}{\sqrt{u^2}}
{{S}^a}_c S^{cb},\label{s2} 
\end{equation}  
 is proportional to $(S^{ab}u_b)^2$ and therefore can be set to zero, given that it does not affect the EOM since it produces a correction which is proportional to the SSC itself. Therefore, we have a choice: we can use either (\ref{s2nc}) or (\ref{s2}),  as they lead to the same result.  In what follows we will use (\ref{s2}), although we will also provide the equivalent result using (\ref{s2nc}) in appendix A.\\

The operator in (\ref{s2}) reproduces the well known LO spin quadrupole contribution to the potential
\beq
\label{ss2pn}
V_{2PN}^{s^2} = -C^{(1)}_{ES^2}\frac{m_2}{2m_1r^3} \left({\bf S}_1\cdot{\bf S}_1 - 3{\bf S}_1\cdot{\bf n}{\bf S}_1\cdot{\bf n}\right) + 1\to 2 .
\eeq

In the case of a rotating black hole $C_{ES^2}=1$, and this term represents the
non--vanishing quadrupole moment of the Kerr solution. The coefficient for other 
compact objects can be calculated via  a matching procedure.
In this paper we will compute the corrections from (\ref{s2}) to the gravitational potential at 3PN. As we will show below there are no other contributions at 3PN from other higher dimensional operators.\\

By using the EFT power counting rules it is easy to organize the perturbative expansion in a systematic way. To obtain Post-Newtonian corrections one calculates ${\cal R}$, or the effective potential, perturbatively, without imposing the SSC (see \cite{ss} for details). The advantage of this approach is that one does not have to worry about complicated algebraic structures. The price to pay is the need of a spin tensor rather than a three vector, though once we find the EOM via (\ref{eomV}) we may write our results in terms of three vector and the coordinate velocity. For the former a precession equation can be obtained.
 
\section{Next to Leading Order $S^2$ potentials}

The algorithm to calculate potentials in the EFT approach is quite simple. First of all we take all of 
the terms in the Routhian and collects them according to their order in the power counting (see \cite{nrgr,nrgr3} for details of the power counting).  Then we draw all possible
Feynman diagrams at the order of interest. 
Each diagram is written in terms of a set of scalar integrals, and the diagram adds a term to the effective action given by $-i \int dt V$, where $V$ is the contribution of that diagram to the effective potential (see \cite{ss}). Throughout this section we will suppress the factors of ``$\int dt$"  in the diagrams. See \cite{TASI} for details on EFTs.

\subsection{ Feynman rules: spin--graviton vertex}

In the weak gravity limit  the relevant spin couplings are \cite{nrgr3,eih,ss}
\begin{eqnarray}
L^{NRGR}_{1PN} &=& \frac{1}{2m_p}H_{i0,k}S^{ik},\label{sgnr1}\\
L^{NRGR}_{1.5PN} &=& \frac{1}{2m_p}\left(H_{ij,k}S^{ik}u^j + H_{00,k}S^{0k}\right),\label{sgnr15}\\
L^{NRGR}_{2PN} &=& \frac{1}{2m_p}\left(H_{0j,k}S^{0k}u^j +
H_{i0,0}S^{i0}\right)+ \frac{1}{4m^2_p}S^{ij}\left(H^{\lambda}_j H_{0\lambda,i} -
H^k_j H_{0i,k}\right).\label{sgnr2}
\end{eqnarray}
We refer the reader to \cite{nrgr1} for all the other Feynman rules not involving spin.

\subsection{$S^2$ terms from non--linear gravitational effects}

\begin{figure}[h!]
    \centering
    \includegraphics[width=7cm]{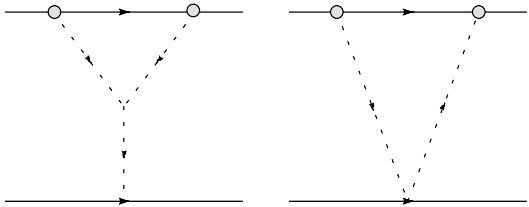}
\caption[1]{Non--linear $S^2$ terms. The blob represents a spin insertion.}\label{nonl}
\end{figure}

The Feynman rules in (\ref{sgnr1},\ref{sgnr15},\ref{sgnr2}) will contribute to the potentials terms which do not arise from either finite size
effects or the Routhian term of the form $RSS$. These non-linear gravitational terms contribute from the diagrams shown in (\ref{nonl}).

It is easy to show that the ``seagull" diagram (the one not containing the three graviton interaction) 
contributes at 4PN, since $\langle H_{00}H_{0i}\rangle=0$.
(Where bracketed polynomials correspond to time ordered products and reduce down to
retarded Greens function for quadratics.) Therefore, for the subleading non--linear $S^2$ effects the only non-vanishing contribution comes from the  three graviton vertex which resides in  the three point function, $\langle H_{00}H_{0i}H_{0j}\rangle$. This contraction can be easily handled by a short  Mathematica routine \footnote{In \cite{kol2} it has been shown that this diagram can be eliminated by a different choice of the metric variables.} which  can be found at \cite{workshop}. The result from the non--linear effects at 3PN is
\beq
\label{nonls2}
\mbox{Fig.~\ref{nonl}} = i\frac{m_2 G_N^2}{2r^4}S^{ik}_1S^{jk}_1(\delta^{ij}-2n^in^j)=          
 i\frac{m_2 G_N^2}{r^4} ({\bf S}_1\cdot {\bf n})^2,
\eeq
from which we obtain the potential
\beq
\label{potnl}
V^{s^2}_{nl} =- \frac{m_2 G_N^2}{r^4} ({\bf S}_1\cdot {\bf n})^2.
\eeq

\subsection{$S^2$ terms from the Routhian}

  Recall that the Routhain includes a terms quadratic in the spin given by
\begin{equation}
\label{lrs2}
L_{RS^2} = -\frac{1}{2m_q}R_{de a b}(x_q)S^{c d}_{q} S^{a b}_{q} \frac{u^e_q u_{qc}}{\sqrt{u^2}},
\end{equation}
whose LO contribution shows up at 2.5PN, 
\begin{equation}
L^{2.5PN}_{RS^2} =-\frac{1}{2m} R_{i0jk}S^{jk}S^{ci}v_c \sim \frac{1}{2m}H_{0k,ij}S^{jk}S^{ci}v_c. 
\end{equation}

To get a 3PN net effect we need to calculate a diagram similar to Fig. \ref{fs1n}b, with the box representing now an insertion of $L_{RS^2}$, to produce a contribution
\beq
\langle L^{2.5PN}_{RS^2} (m_2 H_{0i}v_2^i)\rangle = i \frac{2m_2G_N}{m_1r^3} \left[ 3{\bf n}\cdot({\bf v}_2\times {\bf S}_1)n^l -({\bf v}_2\times {\bf S}_1)^l\right] S_1^{cl}v_c. 
\eeq

Also at 3PN we have a contribution from (\ref{lrs2}),
\begin{equation}
L^{3PN}_{RS^2} = \frac{1}{4m}\left(2H_{li,0j}S^{ij}+2H_{00,lj} S^{j0} - S^{ij}v^k (H_{lj,ki}+H_{ki,lj} -H_{li,kj}-H_{kj,li}) \right)S^{cl}v_c\label{rs23pn},
\end{equation}
which contracts with a LO mass insertion to account for a 3PN contribution given by,
\bea
\langle L^{3PN}_{RS^2} (\frac{m_2}{2} H_{00})\rangle &=& -i \frac{m_2G_N}{m_1r^3}
\left[ \left({\bf v}_2-3{\bf v}_1)\times {\bf S}_1\right)^l + 6 n^l ({\bf v}_1\times{\bf S}_1)\cdot{\bf n} + 3 ({\bf n}\times {\bf S}_1)^l ({\bf n}\cdot{\bf v})\right] S_1^{cl}v_c.\nn \\ 
\eea
Adding both pieces together we end up with the contibution to the potential,
\beq
\label{vrs2}
V^{RS^2}_{3PN} = \left({\bf \tilde a}^{so}_{1(1)}\right)^l S_1^{cl}v_{1c} + 1 \to 2,
\eeq
with ${\bf \tilde a}^{so}_{1(1)}$ the ${\bf S}_1$ piece of the acceleration in the local frame,
\begin{equation}
\label{covasoloc}
{\bf \tilde a}^{so}_{1(1)} =  \frac{m_2G_N}{m_1r^3} \left[ -3{\bf v} \times {\bf S}_1 + 6{\bf n} ({\bf v} \times {\bf  S}_1)\cdot {\bf n} + 3 {\bf n}\cdot{\bf v} ({\bf n} \times {\bf  S}_1) \right]. 
\end{equation}

Notice that we substituted the SSC for the $S_1^{j0}$ term inside the bracket in (\ref{rs23pn}) since it multiplies the SSC itself. From here we can use (\ref{eomV}) to obtain the EOM from which we get
\begin{equation}
\label{eomspin}
\frac{d{\bf S}_1}{dt} = \ldots + ({\bf \tilde a}^{so}_{1(1)}\times {\bf S}_1)\times{ \bf v}_1 +\ldots
\end{equation}

A similar expression straightforwardly follows by using the acceleration dependent form of the extra piece in the Routhian after the redefinition in (\ref{dsu}).

\subsection{Subleading $S^2$ terms from finite size effects}

Let us now consider the finite size corrections. Using the power counting \cite{nrgr1} it is straightforward to show that the LO contribution from  (\ref{s2}) scales $L_{ES^2} \sim \sqrt{L}v^4$, for maximally rotating compact objects, with $C_{ES^2} \sim 1$. 
As we mentioned before, this term generates a gravitational potential of the form of (\ref{ss2pn}) at 2PN.
As it is well known, rotating BHs or NSs have a quadrupole moments given by $Q_{bh}=-aS^2/m$ ($G=c=1$), with $m,S$, the mass and spin respectively \cite{mtw}. For a BH we have $a=1$, 
and for NS $a$ ranges between $4$ and $8$ depending on the equation of state of the neutron star matter \cite{poisson2}. Matching this result with the effective theory we find $C_{ES^2} \equiv a$, which is consistent with what we expect from naturalness
arguments. It also tells us that this contribution is enhanced for NSs with respect to BHs, since the latter seems to provide a lower bound for $C_{ES^2}$ in (\ref{s2})
\footnote{Possibly such a bound could follow from studying graviton scattering off the finite sized object using the dispersion relation techniques developed in \cite{prl4}.}.
 
In order to calculate the 3PN NLO correction due to the finite size of the objects we need to take care of two things: first we need to expand (\ref{s2}) in powers of the relative velocity
$v$, and compute all possible $S^2$--orbit diagrams such that the
net scaling goes like $v^6$; and second we have to make sure we are not missing any other higher dimensional operator which could contribute at 3PN order. 

\subsubsection{All Possible Higher dimensional operators}

Let us start with the higher dimensional operators. At NLO we have a few operators  that could contribute. Let us start using ${\cal S}^{ab}$ to begin with since its use  guarantees the preservation of the SSC upon evolution.
Furthermore, let us put aside reparameterization invariance (RPI) for the time being . If we define ${\cal S}^a=\epsilon^{abcd}{\cal S}_{bc}u_d$ we can write down the following new terms in the action
\begin{eqnarray}
& & D_a B_{cd}{\cal S}^{ac}{\cal S}^d, \\ & &  D_a E_{cd}{\cal S}^{ac}{\cal S}^d .
\end{eqnarray}
where $B_{ab}$ represents the magnetic part of the Weyl tensor.

These terms are self--induced effects which can be generated by diagrams in the one--point function with two spin insertions.  These operators scale as $\sqrt{L}v^6$ and $\sqrt{L}v^7$ respectively. Both will contribute beyond 3PN since the magnetic component of the Weyl tensor does not couple to a LO mass insertion, e.g. $\langle B_{ab} H_{00}\rangle =0$. We may wonder  about operators with only two derivatives since we could also have
\begin{eqnarray}
& &  E_{bc}{\cal S}^{ab}u_a {\cal S}^{cd}u_d\\ & & B_{cb}{\cal S}^{ac}{\cal S}^b u_a \label{second}.
\end{eqnarray}

Notice however that these operator cannot contribute since ${\cal S}^{ab}u_b \equiv 0$, identically. 
We could, however, have chosen to write these operators in terms of $S^{ab}$, in which case  the first one does not contribute, since it is proportional to $(S^{ab}u_b)^2$ and thus  always has a vanishing contribution to the EOM. On the other hand, the second term in (\ref{second}) would be equivalent to
\beq
R_{abcd}S^{bc} S^{de}u^a u_e,
\eeq
and we immediately recognize this is our extra term in the Routhian of (\ref{actR}). 
Notice that, once the SSC is enforced, this term does not contribute to the $n$--point function. 
This makes fixing its coefficient, by matching to the full theory, ambiguos.  Which is to say that
the coefficient is independent of the underlying theory and must be fixed algebraically.
This is consistent with the fact that Wilson coefficient is  fixed by the consistency of the SSC, in this case the covariant choice, as discussed in the previous section.

Finally we should consider the term 
\beq
L_{BS^2}=C_{BS^2} B_{cd} S^{cb}S_b^d,
\eeq
similar to the electric quadrupole. The contribution of this operator to the potential stems from the coupling to  an ${\cal O}(v)$ mass insertion of the form  $H_{0i}v^i$. The LO potential at 2.5PN reads
\beq
V^{s^2}_{BS^2} = C_{BS^2}\frac{6G_N}{r^3} ({\bf n}\times{\bf v}_2)\cdot{\bf S}_1({\bf n}\cdot{\bf S}_1) + 1 \to 2.
\eeq

It turns out however that this term has a vanishing Wilson coefficient, e.g. $C_{BS^2}=0$, due to parity conservation. For instance we can easily show that it vanishes for the case of a Kerr BH by comparison with the multipole expansion of the Kerr metric in harmonic coordinates\footnote{There is no $S^2$ term in the expansion of the $h_{0i}$ component of the Kerr metric \cite{thorne}.}.

\subsubsection{ NLO corrections to the (spin$^2$)quadrupole--monopole interaction induced by $L_{ES^2}$.}

\begin{figure}[h!]
    \centering
    \includegraphics[width=8cm]{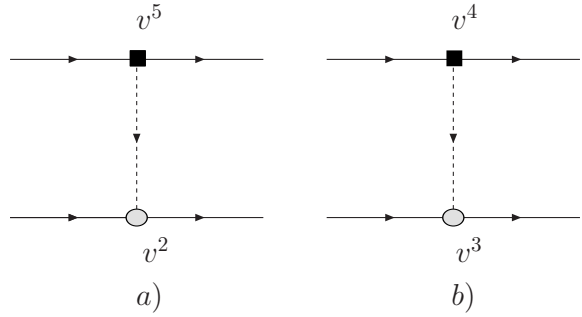}
\caption[1]{Leading order finite size spin$^2$--spin
contributions. The black square represents an insertion of the finite size operator}\label{fs3}
\end{figure}

We are then left to compute the subleading corrections due to (\ref{s2}).
Using the power counting rules of NRGR for spinning bodies \cite{nrgr3} we obtain the following new vertices
\begin{eqnarray}
L_{ES^2}^{2PN} &=& -\frac{C_{ES^2}}{4m m_p}H_{00,ij}S^{ik}S^{jk},\label{fsr1}\\
L_{ES^2}^{2.5PN} &=& -\frac{C_{ES^2}}{2m m_p} H_{0l,ij}v^l S^{ik}S^{jk},\label{fsr2}\\
L_{ES^2}^{3PN} &=& \frac{C_{ES^2}}{2m
m_p}\left[\frac{1}{2}H_{00,ij}S^{i0}S^{j0}+
 S^{ik} S^{jk}\left(H_{il,0j}v^l-\frac{1}{2}H_{lr,ij}v^rv^l + H_{li,jr}v^lv^r\right.\right.\\
& &  \left.\left.  - 
\frac{\vec{v}^2}{4}H_{00,ij}+\frac{1}{2} H_{ij,l}a^l\right)+ S^{0k} S^{jk}H_{00,lj}v^l\right]\nonumber\\
& & + \frac{C_{ES^2}}{8m
m^2_p} S^{ik} S^{jk}\left[H_{00,i}H_{00,j}+H_{0i,l}H_{0l,j}-H_{0l,j}H_{0l,i}
\right.\nonumber\\
& & +\left. H_{00,l}(H_{ij,l}-H_{il,j}-H_{jl,i}) +
H_{0l,i}H_{0j,l}-H_{i0,l}H_{j0,l}\right.\nonumber\\
& & + \left. H_{00}H_{00,ij} - 2 H_{li}H_{00,lj}\right]\nonumber \label{fsr3}
\end{eqnarray}
where $a^l \equiv \frac{d v^l}{dt}$ is the particle's acceleration\footnote{This term follows from integrating by parts the piece proportional to $-\frac{C_{ES^2}}{4mm_p}S^{ik}S^{jk}h_{ij,\alpha\beta} v^\alpha v^\beta$ in the expansion of the electric component of the Weyl tensor in (\ref{s2}). This term was incorrectly discarded in our previous code as a total derivative. We thank A. Ross for pointing this out to us.}, and we have discarded terms which are total derivatives at 3PN. At this order we must also include diagrams with double graviton
exchange. The quadratic term in (\ref{fsr3}) arises from expanding
$\frac{E_{\mu\nu} e^\mu_a e^\nu_b }{\sqrt{u^2}}$ to second order in the metric and tetrad
perturbation. 
Notice that  the LO finite size effects in the spin$^2$--spin sector  contributions, shown in Fig. \ref{fs3}, start at 3.5PN and therefore can be ignored.

\begin{figure}[h!]
    \centering
    \includegraphics[width=8cm]{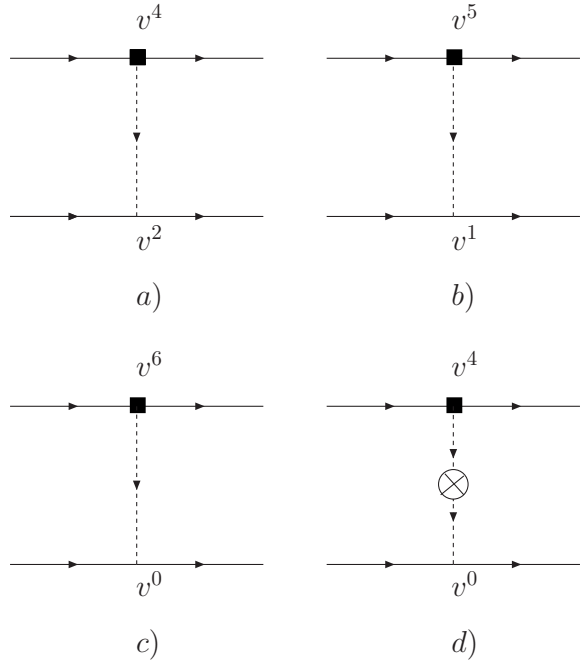}
\caption[1]{Diagrams contributing to the 3PN  which do not involve non-linearities. The cross corresponds to a
propagator correction, i.e. a retardation effect.}\label{fs1n}
\end{figure}
\begin{figure}[h!]
    \centering
    \includegraphics[width=9cm]{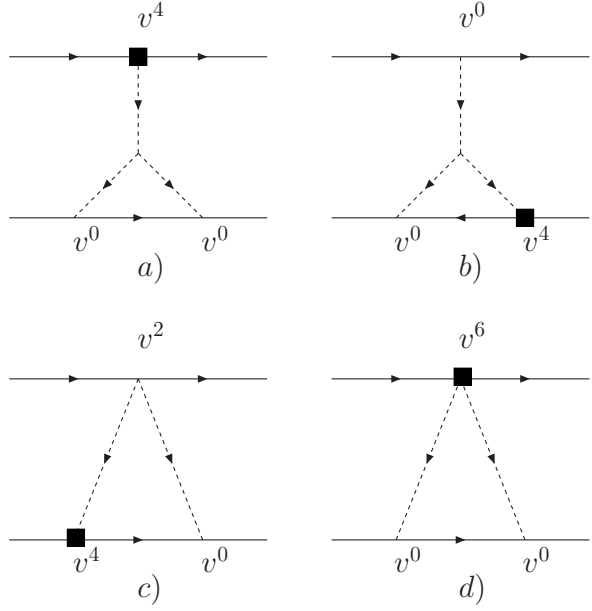}
\caption[1]{Non--linear finite size contributions to the 3PN
spin--orbit potential.}\label{fs2n}
\end{figure}

Fig. \ref{fs1n} shows 3PN contribution from one graviton exchange, with the last diagram
corresponding to the first retardation effect, while Fig. \ref{fs2n}  gives the contributions form non-linearities. 
Calculating these diagrams is straightforward. The results for each one of the diagrams are
\begin{eqnarray}
& & { } \mbox{Fig.~\ref{fs1n}a} = \frac{i C^{(1)}_{ES^2}3m_2}{4m_1}\frac{G_N}{r^3}~{\bf v} _2^2 S_1^{ik}S_1^{jk}(3n^in^j-\delta^{ij})\\
& & { } \mbox{Fig.~\ref{fs1n}b} = -\frac{i C^{(1)}_{ES^2}m_2}{m_1}\frac{2G_N}{r^3}~{\bf v}_2\cdot{\bf v} _1 S_1^{ik}S_1^{jk}(3n^in^j-\delta^{ij}),\nonumber\\
& & { } \mbox{Fig.~\ref{fs1n}c} = - \frac{i C^{(1)}_{ES^2}m_2}{m_1}\frac{G_N}{2r^3}\left[  S_1^{i0} S_1^{j0}(3n^in^j-\delta^{ij})-2  S_1^{0k}S_1^{jk}(v_1^j-3{\bf v} _1\cdot{\bf n} n^j)\right.  \nonumber \\   & & { } + \left. 2S_1^{ik}S_1^{jk} \left( (3{\bf n} \cdot{\bf v} _1n^jv_1^i-v_1^iv_1^j)  - \frac{3}{4}{\bf v} _1^2(3n^in^j-\delta^{ij}) + (v_1^iv_2^j-3{\bf v} _2\cdot{\bf n} n^jv_1^i) -\frac{1}{2}\delta^{ij} {\bf a}_1\cdot {\bf r}\right)\right]\nonumber
\end{eqnarray}
for the instantaneous one--graviton exchanges, and
\begin{eqnarray}
& &\mbox{Fig~\ref{fs1n}d} = -i C^{(1)}_{ES^2}\frac{G_Nm_2}{4m_1r^3} S_1^{ki}S_1^{kj}\left[{\bf v} _1\cdot {\bf v} _2(\delta^{ij}-3 n^i n^j) - 3 (v_1^jn^i +v_1^in^j){\bf v} _2\cdot{\bf n}\right.\nonumber\\
& & \left. - 3 {\bf v} _1\cdot{\bf n} (v_2^jn^i + v_2^in^j)+v_1^iv_2^j+v_2^iv_1^j- 3 {\bf v} _1\cdot{\bf n}  {\bf v} _2\cdot{\bf n}  (\delta^{ij}-5n^in^j)\right]
\end{eqnarray}
for the correction stemming from expanding the propagator.\\ 

For the non--linear terms, since the LO finite size insertion is proprtional to $H_{00,ij}$,  we need the three point function $\langle H_{00}H_{00}H_{00}\rangle$, which we can get from the results for the spinless case in \cite{nrgr}, and just correct for the worldline insertions. We end up with
\begin{eqnarray}
& & { } \mbox{Fig~\ref{fs2n}a} = -i C^{(1)}_{ES^2}\frac{G_N^2m^2_2}{m_1r^4} (4n_in_j-\delta_{ij})S_1^{ik}S_1^{jk},\\
& & { } \mbox{Fig~\ref{fs2n}b} = -i C^{(1)}_{ES^2}\frac{G_N^2m_2}{r^4} (3n_in_j-\delta_{ij})S_1^{ik}S_1^{jk},\\
& & { } \mbox{Fig~\ref{fs2n}c} = i C^{(1)}_{ES^2}\frac{G_N^2m_2}{2r^4} (3n_in_j-\delta_{ij})S_1^{ik}S_1^{jk},\\
& & { } \mbox{Fig~\ref{fs2n}d} = - i C^{(1)}_{ES^2}\frac{G_N^2m^2_2}{m_1r^4} ( 2n_in_j-\delta_{ij})S_1^{ik}S_1^{jk}.
\end{eqnarray}

As in \cite{ss}, we split  the potential at 3PN into  two pieces

\bea
\label{bfs0}
V^{s^2}_{S^{i0}} &=& C^{(1)}_{ES^2}\frac{G_Nm_2}{2m_1r^3}\left[ S_1^{j0}S_1^{i0}(3n^in^j-\delta^{ij}) 
-2 S_1^{k0} \left( ({\bf v}_1\times{\bf S}_1)^k - 3 ({\bf n}\cdot {\bf v}_1)({\bf n}\times{\bf S}_1)^k\right)\right] \nn \\ & & + 1 \to 2 
\eea
and
\bea
\label{bfs}
V^{s^2}_{\bf S} &=& C^{(1)}_{ES^2}\frac{G_Nm_2}{2m_1r^3}\left[ {\bf S}_1^2 \left( 6 ({\bf n}\cdot{\bf v}_1)^2 - \frac{15}{2} {\bf n}\cdot{\bf v}_1{\bf n}\cdot{\bf v}_2 + \frac{13}{2}{\bf v}_1\cdot{\bf v}_2 - \frac{3}{2}{\bf v}_2^2 - \frac{7}{2}{\bf v}_1^2-2{\bf a}_1\cdot {\bf r} \right)  \right. \nn \\ & +&   ({\bf S}_1\cdot {\bf n})^2 \left ( \frac{9}{2}({\bf v}_1^2+{\bf v}_2^2)-\frac{21}{2}{\bf v}_1\cdot{\bf v}_2 - \frac{15}{2} {\bf n}\cdot{\bf v}_1 {\bf n}\cdot{\bf v}_2\right)+
2{\bf v}_1\cdot{\bf S}_1{\bf v}_1\cdot{\bf S}_1\nn \\ &-& \left. 3{\bf v}_1\cdot{\bf S}_1{\bf v}_2\cdot{\bf S}_1 
-6 {\bf n}\cdot{\bf v}_1{\bf n}\cdot{\bf S}_1{\bf v}_1\cdot{\bf S}_1+ 9 {\bf n}\cdot{\bf v}_2{\bf n}\cdot{\bf S}_1{\bf v}_1\cdot{\bf S}_1+ 3 {\bf n}\cdot{\bf v}_1{\bf n}\cdot{\bf S}_1{\bf v}_2\cdot{\bf S}_1\right]\nn \\
& & + C^{(1)}_{ES^2}\frac{m_2G_N^2}{2r^4}\left(1+\frac{4m_2}{m_1}\right) \left( {\bf S}_1^2 - 3({\bf S}_1\cdot{\bf n})^2\right) + 1 \to 2 . 
\eea


\subsection{The $S^2$ potential in the covariant SSC}

Collecting all the pieces together we end up with the following expression for the spin$^2$ potential to 3PN in the covariant SSC
\bea
\label{vs2covn}
V_{3PN}^{s^2}&=& V^{RS^2}_{3PN}+V^{s^2}_{S^{0i}}+V^{s^2}_{\bf S}+V^{s^2}_{nl}+V^{s^2}_{2PN}
\\ &=& C^{(1)}_{ES^2}\frac{G_Nm_2}{2m_1r^3}\left[ S_1^{j0}S_1^{i0}(3n^in^j-\delta^{ij}) 
-2 S_1^{k0} \left( ({\bf v}_1\times{\bf S}_1)^k - 3 ({\bf n}\cdot {\bf v}_1)({\bf n}\times{\bf S}_1)^k\right)\right] 
 \nn \\ & &+ C^{(1)}_{ES^2}\frac{G_Nm_2}{2m_1r^3}\left[ {\bf S}_1^2 \left( 6 ({\bf n}\cdot{\bf v}_1)^2 - \frac{15}{2} {\bf n}\cdot{\bf v}_1{\bf n}\cdot{\bf v}_2 + \frac{13}{2}{\bf v}_1\cdot{\bf v}_2 - \frac{3}{2}{\bf v}_2^2 - \frac{7}{2}{\bf v}_1^2-2{\bf a}_1\cdot {\bf r}\right)  \right. \nn \\ & &+   ({\bf S}_1\cdot {\bf n})^2 \left ( \frac{9}{2}({\bf v}_1^2+{\bf v}_2^2)-\frac{21}{2}{\bf v}_1\cdot{\bf v}_2 - \frac{15}{2} {\bf n}\cdot{\bf v}_1 {\bf n}\cdot{\bf v}_2\right)+ 2{\bf v}_1\cdot{\bf S}_1{\bf v}_1\cdot{\bf S}_1\nn \\ & & - \left. 3{\bf v}_1\cdot{\bf S}_1{\bf v}_2\cdot{\bf S}_1 
-6 {\bf n}\cdot{\bf v}_1{\bf n}\cdot{\bf S}_1{\bf v}_1\cdot{\bf S}_1+ 9 {\bf n}\cdot{\bf v}_2{\bf n}\cdot{\bf S}_1{\bf v}_1\cdot{\bf S}_1+ 3 {\bf n}\cdot{\bf v}_1{\bf n}\cdot{\bf S}_1{\bf v}_2\cdot{\bf S}_1\right]\nn \\
& & -C^{(1)}_{ES^2}\frac{m_2G_N}{2m_1r^3} \left( {\bf S}_1^2 - 3({\bf S}_1\cdot {\bf n})^2 \right)+  C^{(1)}_{ES^2}\frac{m_2G_N^2}{2r^4}\left(1+\frac{4m_2}{m_1}\right) \left( {\bf S}_1^2 - 3({\bf S}_1\cdot{\bf n})^2\right)  \nn \\ & & -\frac{G_N^2m_2}{r^4}\left({\bf S}_1\cdot {\bf n}\right)^2 + \left({\bf \tilde a}^{so}_{1(1)}\right)^l S_1^{0l} + {\bf v}_1\times{\bf S}_1\cdot {\bf \tilde a}^{so}_{1(1)} + 1 \leftrightarrow 2. \nn
\eea

In appendix B we provide a cross check for this potential by taking the extreme mass ratio limit and showing that we reproduce the motion of a test particle in a Kerr background as expected.

\subsection{Divergences and Regularization}

\begin{figure}[h!]
    \centering
    \includegraphics[width=7cm]{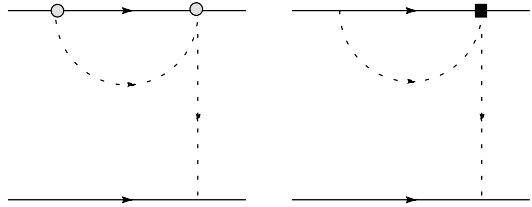}
\caption[1]{Divergent diagrams which renormalize the mass and quadrupole couplings.}\label{div}
\end{figure}

In computing the Feynman diagrams leading to the expression in (\ref{vs2covn}) we encounter divergences of many sorts. First of all from Wick contractions such as the ones represented in Fig.~\ref{div}. 

These diagrams however simply renormalize the couplings of our theory,  $m_q$ and $C^{(q)}_{ES^2}$ coefficients. The divergences are power like  and are thus  "pure counter--terms" (see \cite{TASI} for a discussion for non-experts). Thus, no renormalization group (RG) flow is present in this case, contrary to the case of logarithmic divergences \cite{nrgr,nrgr3,ss}. In the case of dimensional regularization, these diagrams are automatically  set to zero since they involve scaleless integrals. Other divergences are found for instance in diagrams such as Fig.~\ref{fs2n}ab, or similarly in Fig.~\ref{nonl}. These divergences occur when a factor of momentum squared in the numerator from the three graviton vertex cancels the 
intermediate propagator line connecting with the external worldline. Effectively the resulting diagrams appears as in Fig~\ref{div}. These represent a (pure counterterm) renormalization of the mass and the finite size coefficient $C_{ES^2}$.

At higher order, tidally induced logarithmic divergences will appear. As shown in \cite{nrgr3}, the first tidally induced finite size effects starts out at 5PN for maximally rotating compact objects, from a higher dimensional operators of the form
\beq
\label{cds2}
C_{D^2ES^2}(\mu) D^2 E_{ab}S^a_c S^{cb}.
\eeq

In the expression of (\ref{cds2}) $C_{D^2ES^2}(\mu)$ is a Wilson coefficients which obeys a RG--type of equation,
\beq
\mu \frac{d}{d\mu} C_{D^2ES^2} (\mu) \sim \frac{m}{m_p^4}.
\eeq

Notice that (\ref{cds2}) does not contribute to the one--point function, as it is expected from Birkoff's theorem. That is due to the fact that $D^2 h_{\mu\nu} \sim {\cal O}(h^2)$ on shell, and the linear piece in (\ref{cds2}) can be traded, by a field redefinition, for a term proportional to $h^2$. 
Therefore, (\ref{cds2}) does contribute to the (n+1)--point function and that is how we can match with the full theory using scattering amplitudes.
This result generalizes the effacement of internal structure up to 5PN order\footnote{As it was shown in \cite{nrgr} finite size effects enter at 5PN for spinless bodies via terms proportional to $E^2, B^2$ in the worldline. These terms do not contribute to the one--point function as well and matching can be also achieved via comparison of scattering amplitudes in the EFT and full theory sides.}, modulo self--induced effects computed in this paper, to the case of spinning bodies.

\section{The S$^2$ contribution to the spin dynamics to 3PN in the covariant SSC}

From (\ref{vs2covn}) we can obtain the spin dynamics using (\ref{eomV}). The result reads

\bea
\label{s2neom}
\frac{d{\bf S}_1}{dt} = (\omega_{s^2}^{2PN}+{ \omega}^{3PN}_{s^2}) \times {\bf S}_1+({\bf \tilde a}^{so}_{1(1)}\times {\bf S}_1)\times{ \bf v}_1+ ({\bf v}_1 \times {\bf S}_1)\times {\bf A}_1+ {\bf B}_1 \times {\bf S}_1
\eea
where we included the term from the Routhian in (\ref{eomspin}), and
\beq
\omega_{2PN}^{s^2}=3C^{(1)}_{ES^2}\frac{m_2G_N}{m_1r^3}  {\bf n}({\bf S}_1\cdot{\bf n}).
\eeq

Also
\beq
{\bf A}_1 = {\bf v}_1 \times \omega_{2PN}^{s^2},
\eeq
and
\beq
\label{b1}
{\bf B}_1 = \frac{C^{(1)}_{ES^2}m_2}{2m_1r^3} \left[2{\bf v}_1 ({\bf S}_1\cdot{\bf v}_1)-6{\bf v}_1({\bf n}\cdot{\bf v}_1)({\bf S}_1\cdot{\bf n})\right],
\eeq
\bea
{\omega}_{3PN}^{s^2} &=& C^{(1)}_{ES^2}\frac{G_Nm_2}{2m_1r^3}\left[2({\bf S}_1\cdot {\bf n}){\bf n} \left ( \frac{9}{2}({\bf v}_1^2+{\bf v}_2^2)-\frac{21}{2}{\bf v}_1\cdot{\bf v}_2 - \frac{15}{2} {\bf n}\cdot{\bf v}_1 {\bf n}\cdot{\bf v}_2\right)+
4{\bf v}_1 ({\bf v}_1\cdot{\bf S}_1)\right.  \nn \\ &-&  \left. 3{\bf v}_1({\bf v}_2\cdot{\bf S}_1) - 3{\bf v}_2({\bf v}_1\cdot{\bf S}_1)  - 6 {\bf n} ({\bf n}\cdot{\bf v}_1)({\bf v}_1\cdot{\bf S}_1)-6{\bf v}_1 ({\bf n}\cdot{\bf v}_1)({\bf n}\cdot{\bf S}_1) \nn \right.\\ &+& \left. 9 {\bf n}({\bf n}\cdot{\bf v}_2) ({\bf v}_1\cdot{\bf S}_1) +9{\bf v}_1({\bf n}\cdot{\bf v}_2) ({\bf n}\cdot{\bf S}_1) + 3 {\bf n}({\bf n}\cdot{\bf v}_1)({\bf v}_2\cdot{\bf S}_1)\right.\nn 
\\ &+&  \left.  3 {\bf v_2}({\bf n}\cdot{\bf v}_1)({\bf n}\cdot{\bf S}_1)\right]
- C^{(1)}_{ES^2}\frac{m_2G_N^2}{r^4}\left(1+\frac{4m_2}{m_1}\right) 3{\bf n}({\bf S}_1\cdot{\bf n})- \frac{2m_2G_N^2}{r^4} {\bf n}({\bf S}_1\cdot{\bf n})\label{oms2}.
\eea

\subsection{The precession equation}

The spin EOM in (\ref{s2neom}) can be transformed into a precession form by performing the transformation to the NW SSC  which we applied in \cite{ss}
\begin{equation}
{\bf S}^{nw} _1 = \left(1-\frac{1}{2}{\bf \tilde v} _1^2\right){\bf S} _1 + \frac{1}{2}{\bf \tilde v} _1({\bf \tilde v} _1\cdot{\bf S} _1)  + \frac{G_N}{2r^2}\left[ ({\bf n} \times {\bf S}_2)\times {\bf v}_2\right]\times{\bf S}_1 + \ldots
\label{pnshift}
\end{equation}
with $\bf \tilde v$ the velocity in the local frame given by
\beq
{\bf \tilde v}_1 = \left(1+\frac{G_Nm_2}{r}\right) {\bf v}_1 + \frac{G_N}{r^2} {\bf n}\times {\bf S}_2 +\ldots
\eeq
and
\beq
{\bf x}_q \to {\bf x}_q - \frac{1}{2m_q}({\bf v}_q\times {\bf S}_q) + \ldots , \label{pnshift2}
\eeq
where the ellipses represent higher order corrections.

The EOM in terms of ${\bf S}^{nw}$ takes a precession form. This form was already shown for the ${\bf S}^{nw}_1{\bf S}^{nw}_2$ sector \cite{ss}, and now for ${\left({\bf S}^{nw}_1\right)}^2$ effects we have 
\beq
\frac{d}{dt}{\bf S}_1^{nw} = \omega^{nw}_{s^2}\times {\bf S}^{nw}_1
\eeq
with (we suppressed the $nw$ label in the spin vector for simplicity)
\beq
\omega^{nw}_{s^2} = \omega_{s^2}^{2PN}+{\omega}^{3PN}_{s^2} + {\bf B}_1+\frac{1}{2}\left({\bf \tilde a}_{1(1)}+\omega_{s^2}^{2PN}\times{\bf v}_1\right)\times{\bf v}_1 + \delta \omega^{s^2}_{so} + \delta \omega^{2PN}_{s^2}
\eeq
where ${\bf \tilde a}_{1(1)}$ and ${\bf B}_1$ are given by (\ref{covasoloc}) and (\ref{b1}) respectively. We also have the ${\cal O}({\bf S}_1)$ pieces of the shift in the spin--orbit frequency after (\ref{pnshift}) (which effectively takes the covariant result into the NW spin--orbit precession as shown in \cite{ss}) and hence from (\ref{pnshift2}) (see also eq. (74) in \cite{ss}), 
\begin{eqnarray} \delta\omega^{s^2}_{so} &=&  \frac{m_2G_N}{2m_1r^3}\left\{{\bf n}\times \left(\frac{9}{2}{\bf v}_1-6{\bf v}_2\right)({\bf n}\times {\bf v}_1)\cdot {\bf  S}_1 + {\bf v}_1 {\bf S}_1\cdot \left(\frac{3}{2} {\bf v}_1-2{\bf v}_2\right) \right\} \nn \\ & & + \frac{3}{4}\frac{G_N^2m_2^2}{m_1r^4} {\bf n}({\bf n}\cdot {\bf S}_1).
\label{dw1}
\end{eqnarray}
Also from the inverse of (\ref{pnshift}) in the LO finite size term
\begin{equation}
\delta \omega^{2PN}_{s^2} = C_{ES^2}^{(1)}\frac{3G_Nm_2}{2m_1r^3}\left[
{\bf n} ({\bf n}\cdot {\bf S}_1) {\bf v}_1^2 -  {\bf n} ({\bf n}\cdot {\bf v}_1) ({\bf v}_1\cdot {\bf S}_1)\right] \label{deltaw0}.
\end{equation}

\section{The spin potential to 3PN in the covariant SSC}

For completeness here we re--write the spin--spin and spin$^2$ potentials, including also the LO spin--orbit term, 
\bea
\label{fullspin}
V^{spin} &=& \frac{G_Nm_2}{r^2}n^j\left(S^{j0}_1+S^{jk}_1(v_1^k-2v^k_2)\right) -\frac{G_Nm_1}{r^2}n^j\left(S^{j0}_2+S^{jk}_2(v_2^k-2v^k_1)\right) \\
& & -\frac{G_N}{r^3}\left[(\delta^{ij}-3n^i n^j)\left( S^{i0}_1S^{j0}_2+\frac{1}{2}{\bf v}_1 \cdot {\bf v}_2 
S^{ik}_1S^{jk}_2+v_1^mv_2^k S^{ik}_1S^{jm}_2- v_1^k v_2^m S^{ik}_1S^{jm}_2 \right. \right. \nn \\ 
 & & +  \left.
S^{i0}_1S^{jk}_2(v_2^k-v_1^k)+S^{ik}_1S^{j0}_2(v_1^k-v_2^k)\right)+ \frac{1}{2}S_1^{ki}S_2^{kj}\left( 3 { \bf v} _1\cdot{\bf n}  { \bf v} _2\cdot{\bf n}  (\delta^{ij}-5n^in^j) \right. \nn\\   & & + \left. 
   3 {\bf v} _1\cdot{\bf n} (v_2^jn^i+v_2^in^j)+3{\bf v} _2\cdot{ \bf n} (v_1^jn^i +v_1^in^j) -v_1^iv_2^j-v_2^iv_1^j\right) \nn \\ 
 & & + \left. (3n^l{\bf v}_2 \cdot{\bf n} -v_2^l) S_1^{0k}S_2^{kl} + (3n^l{\bf v}_1 \cdot{\bf n} -v_1^l) S_2^{0k}S_1^{kl}  \right] \nn \\ & & +\left(\frac{G_N}{r^3}- \frac{3M G_N^2}{r^4}\right) S_1^{jk}S_2^{ji}(\delta^{ki}-3 n^k n^i)\nn \\
& & + \left\{ C^{(1)}_{ES^2}\frac{G_Nm_2}{2m_1r^3}\left[ S_1^{j0}S_1^{i0}(3n^in^j-\delta^{ij}) 
-2 S_1^{k0} \left( ({\bf v}_1\times{\bf S}_1)^k - 3 ({\bf n}\cdot {\bf v}_1)({\bf n}\times{\bf S}_1)^k\right)\right] 
\right. \nn \\ & &+ C^{(1)}_{ES^2}\frac{G_Nm_2}{2m_1r^3}\left[ {\bf S}_1^2 \left( 6 ({\bf n}\cdot{\bf v}_1)^2 - \frac{15}{2} {\bf n}\cdot{\bf v}_1{\bf n}\cdot{\bf v}_2 + \frac{13}{2}{\bf v}_1\cdot{\bf v}_2 - \frac{3}{2}{\bf v}_2^2 - \frac{7}{2}{\bf v}_1^2-2{\bf a}_1\cdot {\bf r}\right)  \right. \nn \\ & &+   ({\bf S}_1\cdot {\bf n})^2 \left ( \frac{9}{2}({\bf v}_1^2+{\bf v}_2^2)-\frac{21}{2}{\bf v}_1\cdot{\bf v}_2 - \frac{15}{2} {\bf n}\cdot{\bf v}_1 {\bf n}\cdot{\bf v}_2\right)+ 2{\bf v}_1\cdot{\bf S}_1{\bf v}_1\cdot{\bf S}_1\nn \\ & & - \left. 3{\bf v}_1\cdot{\bf S}_1{\bf v}_2\cdot{\bf S}_1 
-6 {\bf n}\cdot{\bf v}_1{\bf n}\cdot{\bf S}_1{\bf v}_1\cdot{\bf S}_1+ 9 {\bf n}\cdot{\bf v}_2{\bf n}\cdot{\bf S}_1{\bf v}_1\cdot{\bf S}_1+ 3 {\bf n}\cdot{\bf v}_1{\bf n}\cdot{\bf S}_1{\bf v}_2\cdot{\bf S}_1\right]\nn \\
& & -C^{(1)}_{ES^2}\frac{m_2G_N}{2m_1r^3} \left( {\bf S}_1^2 - 3({\bf S}_1\cdot {\bf n})^2 \right)+  C^{(1)}_{ES^2}\frac{m_2G_N^2}{2r^4}\left(1+\frac{4m_2}{m_1}\right) \left( {\bf S}_1^2 - 3({\bf S}_1\cdot{\bf n})^2\right)  \nn \\ & &  \left.
-\frac{G_N^2m_2}{r^4}\left({\bf S}_1\cdot {\bf n}\right)^2 + \left({\bf \tilde a}^{so}_{1(1)}\right)^l S_1^{0l} + {\bf v}_1\times{\bf S}_1\cdot {\bf \tilde a}^{so}_{1(1)} + 1 \leftrightarrow 2\right\}, \nn
\eea
which allows us to obtain the ${\cal O}({\bf S}_1{\bf S}_2)$ (computed in \cite{ss}) and ${\cal O}({\bf S}_{1(2)}^2)$ contributions to the EOM to 3PN in the covariant SSC. The latter is imposed after the EOM are obtained via (\ref{eomV}). Note that, to the order we are working in this paper, we can replace the acceleration-dependent term in our Lagrangian by the LO equations of motion, or equivalently integrate it by parts.\\

Missing in (\ref{fullspin}) is the NLO spin--orbit potential. The latter was obtained in \cite{owen,buon}. However, recall that our computation is in terms of the spin in the local frame, $S^{ab}$, whereas in \cite{owen,buon} the spin dynamics was calculated in terms of the spin tensor in the PN frame. e.g. $S^{\mu\nu}$. The result can be translated to the local frame by using the tetrad field $e^a_\mu$ \cite{ss}. The potential in (\ref{fullspin}) therefore completes the computation of spin effects to 3PN. 

\section{conclusions}

In this paper we have completed the calculation of the potential quadratic in spins to 3PN. Together with our previous results in \cite{eih,comment,ss}, the results in this paper complete the computation of the relevant potentials to obtain the spin effects in the EOM to 3PN order.
It is important to note that within NRGR there is no obstacle to go beyond this order using the formalism developed in \cite{nrgr,ss}. In particular, the NNLO corrections should follow with relative ease. The radiation at 3PN from spinning binaries remains to be calculated. The framework
for such a calculation within NRGR was set up in \cite{nrgr}. We report on these effects in \cite{radss}. 

\acknowledgments

This work was supported in part by the Department of Energy contracts DOE-ER-40682-143 and DEAC02-6CH03000. RAP also acknowledges support from the Foundational Questions Institute under grant RPFI-06-18, and funds from the University of California.

\appendix
\section{Using ${\cal S}$ instead of $S$}

In this appendix we compute the subleading finite size corrections using (\ref{s2nc}), which we will explicitly show give rise to identical results to those found in the body of the paper.
The calculation is identical, the only exception is that now we use ${\cal S}^{ab}$ in the diagrams.
As we did in \cite{ss}, we can sum up all the diagrams and hence split the finite size potential at 3PN in two pieces, as we did before in (\ref{bfs}) and (\ref{bfs0}). Hence we have again one term dependent on ${\cal S}^{i0}$,
\bea
\label{bfs0n}
V^{s^2}_{{\cal S}^{i0}} &=& C^{(1)}_{ES^2}\frac{G_Nm_2}{2m_1r^3}\left[ {\cal S}_1^{j0}{\cal S}_1^{i0}(3n^in^j-\delta^{ij}) 
-2 {\cal S}_1^{k0} \left( ({\bf v}_1\times{\bf S}_1)^k - 3 ({\bf n}\cdot {\bf v}_1)({\bf n}\times{\bf S}_1)^k\right)\right] +1 \to 2\nn\\ &=&  C^{(1)}_{ES^2}\frac{G_Nm_2}{2m_1r^3}\left[-3 ({\bf v}_1\times{\bf S}_1)^2+ 3 \left({\bf n}\cdot({\bf v}_1\times{\bf S}_1)\right)^2+6 ({\bf n}\cdot{\bf v}_1)({\bf v}_1\times{\bf S}_1)\cdot({\bf n}\times{\bf S}_1)\right] \nn \\ & & + 1 \to 2  
\eea
where we used (\ref{caso}), and as before at 3PN, 
\bea
V^{s^2}_{\bf S} &=& C^{(1)}_{ES^2}\frac{G_Nm_2}{2m_1r^3}\left[ {\bf S}_1^2 \left( 6 ({\bf n}\cdot{\bf v}_1)^2 - \frac{15}{2} {\bf n}\cdot{\bf v}_1{\bf n}\cdot{\bf v}_2 - \frac{13}{2}{\bf v}_1\cdot{\bf v}_2 - \frac{3}{2}{\bf v}_2^2 - \frac{7}{2}{\bf v}_1^2-2{\bf a}_1\cdot {\bf r}\right)  \right. \nn \\ & +&   ({\bf S}_1\cdot {\bf n})^2 \left ( \frac{9}{2}({\bf v}_1^2+{\bf v}_2^2)-\frac{21}{2}{\bf v}_1\cdot{\bf v}_2 - \frac{15}{2} {\bf n}\cdot{\bf v}_1 {\bf n}\cdot{\bf v}_2\right)+
2{\bf v}_1\cdot{\bf S}_1{\bf v}_1\cdot{\bf S}_1\nn \\ &-& \left. 3{\bf v}_1\cdot{\bf S}_1{\bf v}_2\cdot{\bf S}_1 
-6 {\bf n}\cdot{\bf v}_1{\bf n}\cdot{\bf S}_1{\bf v}_1\cdot{\bf S}_1+ 9 {\bf n}\cdot{\bf v}_2{\bf n}\cdot{\bf S}_1{\bf v}_1\cdot{\bf S}_1+ 3 {\bf n}\cdot{\bf v}_1{\bf n}\cdot{\bf S}_1{\bf v}_2\cdot{\bf S}_1\right]\nn \\
& & + C^{(1)}_{ES^2}\frac{m_2G_N^2}{2r^4}\left(1+\frac{4m_2}{m_1}\right) \left( {\bf S}_1^2 - 3({\bf S}_1\cdot{\bf n})^2\right) + 1 \to 2 . 
\eea

We are still one term short to complete the potential to 3PN. We still need to include the subleading corrections due to (\ref{newsc}) in the LO finite size potential of (\ref{ss2pn}),
\beq
V^{s^2}_{2PN} \to C^{(1)}_{ES^2}\frac{m_2G_N}{2m_1r^3} {\cal S}_1^{ik}{\cal S}_1^{jk}(\delta^{ij}-3n^in^j) + 1\to 2.
\eeq
Equivalently we have
\beq
\label{calspot}
V^{s^2}_{2PN} \to - C^{(1)}_{ES^2}\frac{m_2G_N}{2m_1r^3} \left( {\bf \cal S}_1\cdot {\bf \cal S}_1 - 3{\bf \cal S}_1\cdot {\bf n}
{\bf \cal S}_1\cdot {\bf n}\right) + 1 \to 2,
\eeq
with ${\bf \cal S}_q^{ij} = \epsilon^{ijk}{\cal S}^k$, and 
\beq
\label{algcal}
\{S^i_q,{\cal S}_q^k\} = - \epsilon^{ikl}(S_q^l+({\bf S}_q\cdot{\bf v}_q)v_q^l).
\eeq

The final result for the spin$^2$ potential in the covariant SSC turns out to be the sum of each previous computations in (\ref{vrs2}), (\ref{potnl}), (\ref{bfs0}), (\ref{bfs}) and (\ref{calspot}), resulting in
\bea
\label{vs2cov}
V_{3PN}^{s^2}&=& V^{RS^2}_{3PN}+V^{s^2}_{{\cal S}^{0i}}+V^{s^2}_{\bf S}+V^{s^2}_{nl}+V^{s^2}_{2PN}({\bf \cal S})
\\ &=& C^{(1)}_{ES^2}\frac{G_Nm_2}{2m_1r^3}\left[-3 ({\bf v}_1\times{\bf S}_1)^2+ 3 \left({\bf n}\cdot({\bf v}_1\times{\bf S}_1)\right)^2+6 ({\bf n}\cdot{\bf v}_1)({\bf v}_1\times{\bf S}_1)\cdot({\bf n}\times{\bf S}_1)\right] \nn \\ &+& C^{(1)}_{ES^2}\frac{G_Nm_2}{2m_1r^3}\left[ {\bf S}_1^2 \left( 6 ({\bf n}\cdot{\bf v}_1)^2 - \frac{15}{2} {\bf n}\cdot{\bf v}_1{\bf n}\cdot{\bf v}_2 + \frac{13}{2}{\bf v}_1\cdot{\bf v}_2 - \frac{3}{2}{\bf v}_2^2 - \frac{7}{2}{\bf v}_1^2-2{\bf a}_1\cdot {\bf r}\right)  \right. \nn \\ & +&   ({\bf S}_1\cdot {\bf n})^2 \left ( \frac{9}{2}({\bf v}_1^2+{\bf v}_2^2)-\frac{21}{2}{\bf v}_1\cdot{\bf v}_2 - \frac{15}{2} {\bf n}\cdot{\bf v}_1 {\bf n}\cdot{\bf v}_2\right)+ 2{\bf v}_1\cdot{\bf S}_1{\bf v}_1\cdot{\bf S}_1\nn \\ &-& \left. 3{\bf v}_1\cdot{\bf S}_1{\bf v}_2\cdot{\bf S}_1 
-6 {\bf n}\cdot{\bf v}_1{\bf n}\cdot{\bf S}_1{\bf v}_1\cdot{\bf S}_1+ 9 {\bf n}\cdot{\bf v}_2{\bf n}\cdot{\bf S}_1{\bf v}_1\cdot{\bf S}_1+ 3 {\bf n}\cdot{\bf v}_1{\bf n}\cdot{\bf S}_1{\bf v}_2\cdot{\bf S}_1\right]\nn \\
&+&  C^{(1)}_{ES^2}\frac{m_2G_N^2}{2r^4}\left(1+\frac{4m_2}{m_1}\right) \left( {\bf S}_1^2 - 3({\bf S}_1\cdot{\bf n})^2\right) - \frac{G_N^2m_2}{r^4}\left({\bf S}_1\cdot {\bf n}\right)^2 \nn \\ &-&
C^{(1)}_{ES^2}\frac{m_2G_N}{2m_1r^3} \left( {\bf \cal S}_1\cdot {\bf \cal S}_1 - 3{\bf \cal S}_1\cdot {\bf n} {\bf \cal S}_1\cdot {\bf n}\right) + \left({\bf \tilde a}^{so}_{1(1)}\right)^l S_1^{0l} + {\bf v}_1\times{\bf S}_1\cdot {\bf \tilde a}^{so}_{1(1)} + 1 \to 2 \nn,
\eea
where ${\bf \tilde a}^{so}_{1(1)}$ is given by (\ref{covasoloc}) and ${\cal S}^i = \frac{1}{2}\epsilon^{ijk} {\cal S}^{jk}$, with ${\cal S}^{jk}$ given in (\ref{calsrep}).  with the algebra in (\ref{algcal}) for the ${\bf \cal S}$ pieces. The squared spin contribution to the EOM in the covariant SSC thus read 
\bea
\label{s2eom}
\frac{d{\bf S}_1}{dt} &=& (\omega_{s^2}^{2PN}+{\hat \omega}^{3PN}_{s^2}) \times {\bf S}_1+({\bf \tilde a}^{so}_{1(1)}\times {\bf S}_1)\times{ \bf v}_1 +({\bf S}_1\cdot{\bf v}_1) \omega^{2PN}_{s^2}\times {\bf v}_1,
\eea
where again
\beq
\omega_{2PN}^{s^2}=3C^{(1)}_{ES^2}\frac{m_2G_N}{m_1r^3} {\bf n}({\bf S}_1\cdot{\bf n}),
\eeq
and 
\bea
{\hat \omega}_{3PN}^{s^2} &=& {\bar \omega}_{3PN}^{s^2} + {\omega}_{3PN}^{s^2}\label{omegas}\\
{\bar \omega}_{3PN}^{s^2}&=&3C^{(1)}_{ES^2}\frac{G_Nm_2}{m_1r^3}\left[ {\bf v}_1 ({\bf v}_1\cdot{\bf S}_1)+  {\bf n}\times{\bf v}_1\left({\bf n}\cdot ({\bf v}_1\times {\bf S}_1)\right)\right. \\ & & \left. -{\bf n}({\bf n}\cdot{\bf v}_1)({\bf S}_1\cdot{\bf v}_1) - 
{\bf v}_1({\bf n}\cdot{\bf v}_1)({\bf S}_1\cdot{\bf n})\right] \nn 
\eea
with ${\omega}^{s^2}_{3PN}$ given in (\ref{oms2}). We can now compare the expression in (\ref{s2neom}) with (\ref{s2eom}) and (\ref{omegas}). Using
\beq
\left({\bf B}_1+(\omega^{s^2}_{2PN}\times{\bf v}_1)\times{\bf v}_1\right)\times{\bf S}_1 = {\bar \omega}^{s^2}_{3PN}\times {\bf S}_1= -3C^{(1)}_{ES^2}\frac{G_Nm_2}{m_1r^3} {\bf v}_1^2({\bf n \cdot S}_1)({\bf n \times S}_1)
\eeq
one can show both spin EOM are indeed identical and the equivalence is thus proven.

\section{Checking the extreme mass ratio limit}

In this appendix we compare our results to the extremal limit. That is, we consider the motion of a
test particle with mass $m$ and velocity ${\bf v}$, moving in the background of a spinning BH with mass $M$ and spin ${\bf S}$. In this limit the potential is given by (recall $C_{ES^2}=1$ for a Kerr BH)
\bea
\label{us}
V_{stat}&=&- \frac{G_Nm}{2Mr^3}({\bf S}^2
-3 ({\bf S}\cdot {\bf n})^2) +\frac{G_N^2m}{2r^4}  ({\bf S}^2-3({\bf n}\cdot {\bf S})^2)
-\frac{G_N^2 m}{r^4}({\bf S} \cdot {\bf n})^2
\nn \\ & & -\frac{3G_Nm}{4Mr^3}{\bf v}^2( {\bf S}^2 - 3({\bf S}\cdot {\bf n})^2),
\eea
where we neglect term of order ${\cal O}(m^2)$ (including the acceleration-dependent part of our potential).

Now we would like to compare this to the effective action generated via
\beq
\label{action}
S=-m\int dt \sqrt{g_{\mu \nu} \frac{dx^\mu}{dt} \frac{dx^\nu}{dt}}
\eeq
by using the
harmonic gauge metric \cite{hergt}
\bea
ds^2=g_{00}dt^2+2g_{0\phi}dt d\phi+dl_a^2+dl_{a^2}^2
\eea
where the relevant pieces are (with signature $(-1,1,1,1)$) 
\bea
\label{metric}
g_{00}&=&-1+\frac{2G_NM}{r}-\frac{2G_N^2M^2}{r^2}-\frac{G_NS^2/M-4G_N^3M^3+3G_NS^2/M \cos(2\theta)}{2r^3}\nn \\ & & +\frac{2G_N^2S^2-2G_N^4M^4+4G_N^2 S^2 \cos (2 \theta)}{r^4}\nn \\
g_{0\phi}&=& -\frac{2S G_N \sin^2 \theta}{r}+\frac{2S M G_N^2 \sin^2 \theta}{r^2}+
O(G_N^3,S^3)
\eea
\bea
dl_{a^2}^2=\frac{G_NS^2 }{2 M r}\sin^2 \theta \left(-(1+3 \cos (2 \theta))+\frac{G_NM(\cos (2 \theta)-3)}{r}\right)d\phi^2
\eea
\bea
dl_a^2=((r+G_NM)^2 \sin^2 \theta+O(G_N S)) d\phi^2 .
\eea

For simplicity of comparison, and with no loss of generality, we  consider the motion in a
circular orbit. Notice that we do not expect our results to agree with those in the harmonic gauge
to all orders in $G_N$, since we are working in a ``background field harmonic" gauge. However, one can show that the coordinate transformation to the harmonic gauge will have no effect to the
order we are working in this paper\footnote{Going to the harmonic gauge entails adding a new diagram similar to Fig.~\ref{nonl} with a three graviton interaction that follows from $S_{GF}$ \cite{nrgr}. However, one can show that this extra diagram vanishes and our result for the $S^2$ and spin--spin potentials agree with the harmonic gauge to 3PN. That is not the case for the spinless part of the potential already at 2PN. We thank Andreas Ross for pointing this out to us.}. Indeed plugging in (\ref{metric}) into (\ref{action}) (after changing to our signature convention $(1,-1,-1,-1)$) and expanding leads to the result in (\ref{us}).

\end{document}